\documentclass[twocolumn,pra,twocolumn,superscriptaddress]{revtex4}
\usepackage[latin9]{inputenc}
\usepackage{color}
\usepackage{amsmath}
\usepackage{amssymb}
\usepackage{graphicx}
\usepackage{tikz}
\usepackage{mathrsfs} % cal F fidelity
\usepackage{float}
\usepackage{bbm}
\usepackage{epstopdf}
\usepackage{epsfig}
\usepackage{verbatim}
\usepackage{array}
\usepackage{setspace}
\usepackage{times}

\usepackage[unicode=true,
bookmarks=true,bookmarksnumbered=false,bookmarksopen=false,
breaklinks=false,pdfborder={0 0 1},backref=false,colorlinks=true]
{hyperref}
\hypersetup{
	linkcolor=magenta, urlcolor=blue, citecolor=blue, pdfstartview={FitH}, hyperfootnotes=false, unicode=true}

\newcolumntype{C}[1]{>{\centering\arraybackslash$}p{#1}<{$}}

\begin{document}

\title{High-fidelity geometric gate for silicon-based spin qubits}

\author{Chengxian Zhang}
\author{Tao Chen}
\author{Sai Li}
%\author{Jun-Yi Cao}
\affiliation{Guangdong Provincial Key Laboratory of Quantum Engineering and Quantum Materials, GPETR Center for Quantum Precision Measurement, and School of Physics and Telecommunication Engineering, South China Normal University, Guangzhou 510006, China}
\author{Xin Wang}
\email{x.wang@cityu.edu.hk}
\affiliation{Department of Physics, City University of Hong Kong, Tat Chee Avenue, Kowloon, Hong Kong SAR, China}
\affiliation{City University of Hong Kong Shenzhen Research Institute, Shenzhen, Guangdong 518057, China}
\author{Zheng-Yuan Xue}
\email{zyxue83@163.com}
\affiliation{Guangdong Provincial Key Laboratory of Quantum Engineering and Quantum Materials, GPETR Center for Quantum Precision Measurement, and School of Physics and Telecommunication Engineering, South China Normal University, Guangzhou 510006, China}

\date{\today}

\begin{abstract}

High-fidelity manipulation is the key for the physical  realization of fault-tolerant quantum computation. Here, we present a protocol to realize universal nonadiabatic geometric gates for silicon-based spin qubits. We find that the advantage of geometric gates over dynamical gates depends crucially on the evolution loop for the construction of the geometric phase. Under appropriate evolution loops, both the geometric single-qubit gates and the CNOT gate can outperform their dynamical counterparts for both systematic and detuning noises. We also perform randomized benchmarking using noise amplitudes consistent with experiments in silicon. For the static noise model, the averaged fidelities of geometric gates are around 99.90\% or above, while for the time-dependent $1/f$-type noise, the fidelities are around 99.98\% when only the detuning noise is present. We also show that the improvement in fidelities of the geometric gates over dynamical ones typically increases with the exponent $\alpha$ of the $1/f$ noise, and the ratio can be as high as 4 when $\alpha\approx 3$. Our results suggest that geometric gates with judiciously chosen evolution loops can be a powerful way to realize high-fidelity quantum gates.

\end{abstract}

\maketitle

\section{Introduction}

Spin qubits \cite{Loss.98} in semiconductor quantum dots are promising candidate for physical realization of quantum computation due to their all-electrical control and prospect for scalability \cite{Hanson.07}. %Manipulation of spin qubits has been demonstrated in various materials, including  silicon (Si) \cite{Zwanenburg.13}, gallium arsenide (GaAs) \cite{Petta.05}, and Germanium (Ge) \cite{Watzinger.18} systems.
Silicon-based spin qubits, (SSQ) stands out owing to their relatively long coherence time \cite{Muhonen.14} and high-fidelity gate operations, %\cite{Zajac.18,Takedae.16,Watson.18,Yoneda.18,Veldhorst.14, Veldhorst.15,Huang.19,Xue.19, Brunner.11,Pioro.08,Crippa.19,Yang.19,Kawakami.14,Emerson.19, sigillito.19,Schreiber.18},
which benefits from isotopically enriched $^{28}$Si \cite{Bermeister.14,Zimmerman.14,Veldhorst.14, Veldhorst.15,Takedae.16, Watson.18,Schreiber.18,Chan.18,Yoneda.18,Emerson.19, sigillito.19,Yang.19,Crippa.19,Huang.19} where the nuclear noise is substantially suppressed, as well as techniques operating around certain sweet spots \cite{Shim.16,Russ.16,Zhang.18}. Meanwhile, to further reduce the noises via quantum control, techniques such as dynamical decoupling \cite{Hahn.50}, dynamical corrected gates \cite{xin.12}, %Robert.16,xin.14,
and pulse engineering \cite{Yang.19,Emerson.19} have been put forward. However, the gate fidelity on SSQ still needs to be improved to achieve the stringent requirement set by fault-tolerant quantum computation \cite{Lidar.13}.

%Currently, a quantum gate is achieved on  SSQ via Rabi oscillations between basis states. %$\left|0\rangle\right.$ and $\left|1\rangle\right.$ as shown in Fig.~\ref{fig:qdot_bloch}(c).
%This operation is purely a dynamical evolution and  the acquired dynamical phase is vulnerable to both the systematic  \cite{Zheng.16,tao.18} and detuning \cite{xu.19} noises.
%%Systematic noises stem from the fluctuations in the control field, i.e. the drift of the Rabi frequency which appears in the off-diagonal terms of the evolution Hamiltonian while detuning noises amount to the perturbation in the diagonal terms.
%In SSQ, the detuning noises dominate, which involves both charge and nuclear noises. The charge noises are introduced through spin-orbit coupling \cite{Veldhorst.14,Ferdous.18} and impurities \cite{Hu.06,Culcer.09,Nguyen.11,Chan.18}, and can be suppressed by symmetrically controlling at sweet spots \cite{Reed.16,Martins.16,Bertrand.15}. Even though the nuclear spin noise can be substantially suppressed by isotropic enrichment, residual $^{29}$Si should not be completely ignored \cite{Huang.19}. %To further reduce the noises via quantum control, techniques such as dynamical decoupling \cite{Hahn.50}, dynamical corrected gates \cite{Robert.16,xin.14,xin.12}, and pulse engineering \cite{Yang.19,Emerson.19} have been put forward.
%While these techniques have been vastly successful in compensating errors in the dynamical phases, \underline{no correction on} geometrical phases are made.

In contrast to the dynamical phase for which errors accumulate during the evolution, the geometric phase benefits from its global property
%. Since its discovery by Berry
\cite{Berry.84, Wilczek.84,Aharonov.87}, i.e., it
%became clear that the geometric phase
is determined only by the closed path of the cyclic evolution and is robust against certain types of local noises. This nice feature has inspired  geometric quantum computation (GQC) \cite{ Zanardi.99, Pachos.99}. By using adiabatic cyclic evolution, the geometric gates have been demonstrated \cite{Zanardi.99,Pachos.99,Falci.00,Duan.01,Solinas.03,Wu.13,Toyoda.13,Huang.19geo}. However, the adiabatic limit means that the gates are rather slow, which consequently exposes the qubit to the environment for an overly long time, making it infeasible for realistic quantum computation. To lift the adiabatic limit and speed up the quantum gate, implementation of nonadiabatic geometric gates based on Abelian \cite{Xiang.01,Zhu.02, Wang.16,Zhao.17,tao.18} and non-Abelian \cite{Erik.12,Xu.12} phases have been proposed. The non-Abelian geometric gates have been successfully demonstrated in various systems in experiments \cite{feng.13,abdumalikov.13,Arroyo.14,zu.14,yale.16,Zhou.17,Sekiguchi.17,Li.17,Xu.18,zhu.19,yan.19,egger.19}. However, their realization in quantum dot systems usually requires manipulation of three or more energy levels \cite{Mousolou.14,Mousolou.18}, and is difficult to apply to spin qubits which typically involve two levels.
The Abelian geometric gates have been demonstrated in two-level systems \cite{Zhao.17,tao.18}. In these cases, the geometric gate can be robust against the systematic noise but is more vulnerable to the detuning noise. Usually, the detuning noise is the leading error in many  systems, such as the superconducting circuit and silicon-based quantum dots. Recently, experiments based on the superconducting platform \cite{xu.19} have shown that several individual single-qubit gates can be robust to the detuning noise. Typically, the geometric gate works well under the quasi-static-noise approximation as demonstrated before \cite{tao.18}, where noise is assumed to vary with a much longer time scale compared to that of the gate operation. For the realistic time-dependent noise, one may suppose that the geometric gate is performing effectively for the low-frequency components of the noise. However, whether the geometric gate can still outperform the  dynamical gate at high frequencies remains unclear. On the other hand, the noise spectrum with which the noise is concentrated on for the silicon quantum dot may be different compared to the superconducting circuits. Thus, it is very important to evaluate the geometric gate within a wide range of noise spectrum.

%However, how to generate an arbitrary single-qubit gate that is resilient to detuning noise is still a key challenge, which is also crucial for the construction of the two-qubit geometric gate.

In this work, we present a full theoretical proposal to implement universal GQC in SSQ. We analyze the effect of  systematic noises and detuning noises typical in the silicon system. Comparing the fidelity of the geometric gates and their dynamical counterparts, we find that the robustness of the geometric gates against noises depends crucially on the path taken during the cyclic evolution. By carefully choosing the evolution loop, both detuning noises and systematic noises can be effectively suppressed. %In addition, we also propose how to implement the geometric CNOT gate and discuss how to improve its  performance.
On the other hand, to quantitatively determine the improvement afforded by the geometric gates, we perform randomized benchmarking (RB), which is conducted by comparing the sequences composed of the dynamical and the designed geometric gates within the single-qubit Clifford group. We find that, for the static Gaussian noise,  the averaged fidelities of geometric gates are around 99.90\% or above. Meanwhile, we also consider the time-dependent noise, namely the $1/f$ type whose power spectral density is proportional to $1/\omega^\alpha$ where $\alpha$ indicates the correlation within the noise. We find that both the detuning and systematic noise can be suppressed well. For large $\alpha$, the geometric gates improve the fidelity by a factor of 2 or more over the dynamical ones. Our results suggest that geometric quantum gates can be a powerful alternative to realize high-fidelity quantum manipulations of SSQ.

%The remainder of this work is organized as follows. In Sec. \ref{sec:model} we describe the method to implement both single- and two-qubit geometric gates. In Sec. \ref{sec:comparison} we compare the fidelities between the dynamical and the geometric gates. In Sec. \ref{sec:rb} we show results of RB. In Sec.\ref{sec:conclusion} we conclude.

\begin{figure}
	\includegraphics[width=1\columnwidth]{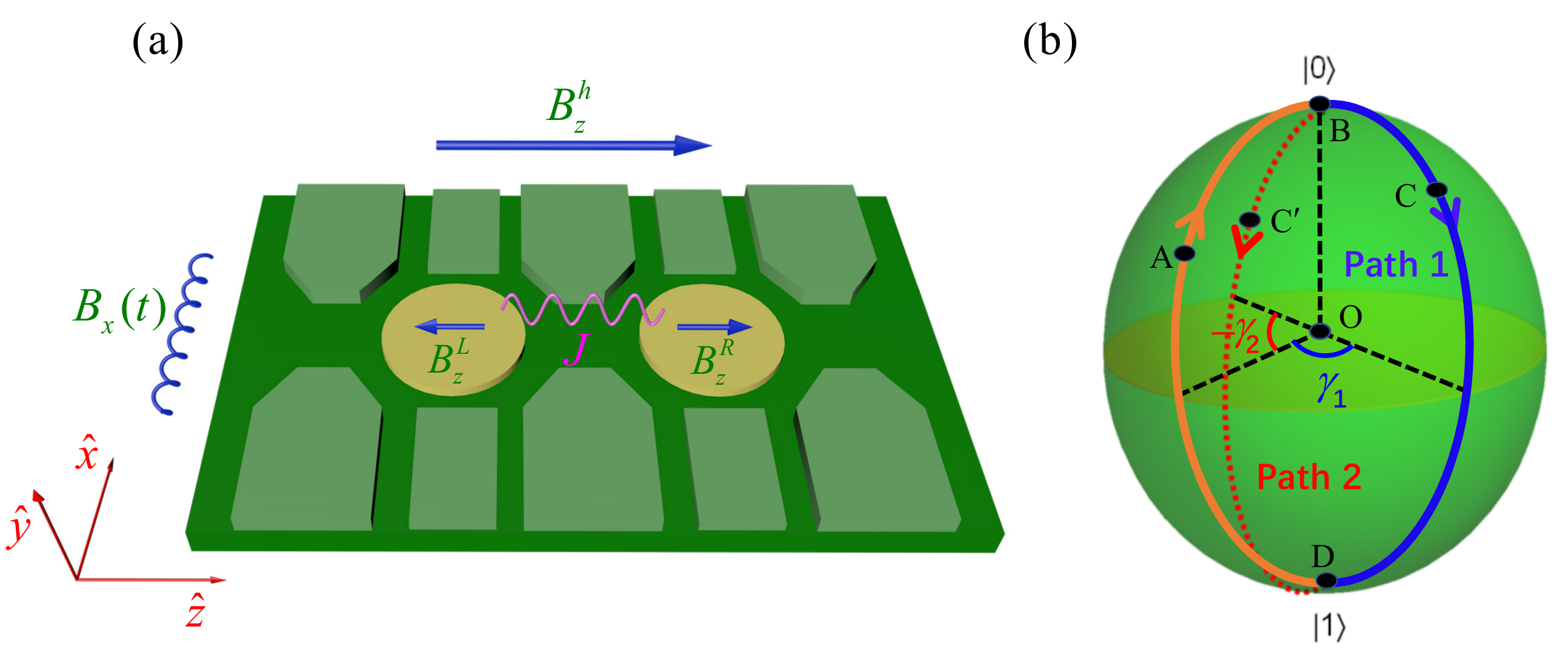}
	\caption{Illustration of the SSQ. (a) Spin qubits in the silicon-based double quantum dot. The single-qubit gate is obtained by operating the magnetic field, while the two-qubit gate is implemented with their Heisenberg interaction. (b) The evolution paths of state $\left|\psi_{+}\right\rangle$. The geometric phase can be achieved through the cyclic evolution along the path $A-B-C-D-A$ and $A-B-C^{'}-D-A$, which correspond to the geometric gates in path 1 and 2, respectively.}
	\label{fig:qdot_bloch}
\end{figure}

\section{Universal nonadiabatic geometric gates}\label{sec:model}
\subsection{Geometric single-qubit gate}
We first show how to implement the nonadiabatic geometric single-qubit gates using a SSQ. As shown in Fig.~\ref{fig:qdot_bloch}(a), the two electron spins are confined in the silicon-based double-quantum-dot system with each electron occupying either the left (L) or right (R) dot. The electron can be either spin up $\left|\uparrow\rangle\right.$ or spin down $\left|\downarrow\rangle\right.$. The basis states for the single qubit are therefore $\left\{\left|0\rangle\right.= \left|\uparrow\rangle\right., \left|1\rangle\right.=\left|\downarrow\rangle\right. \right \}$. Each dot experiences a magnetic field of $B_{L}=(B_{x}^{L}(t),0,B_{z}^{h}+B_{z}^{L})$ or $B_{R}=(B_{x}^{R}(t),0,B_{z}^{h}+B_{z}^{R})$. Here, the magnetic fields are in energy units and we use $\hbar=1$ for convenience. Specifically, $B_{z}^{h}$ denotes the static homogeneous magnetic field in the $z$ direction which can be as large as GHz in the experiment to lift the spin degeneracy \cite{Huang.19}. $B_{z}^{Q}$ ($Q=L, R$) is the local inhomogeneous component to obtain distinct resonance frequencies for an individual qubit. The effective Zeeman splitting for each spin is therefore $E_{z}^{Q}=g\mu_B (B_{z}^{h}+B_{z}^{Q})$. Apart from the static magnetic field,  $B_{x}^{Q}(t)=B_{x}^{Q,0}+B_{x}^{Q,1}\rm{cos}(\omega_{Q}t+\phi)$ denotes the transverse time-dependent oscillating field perpendicular to $B_{z}^{h}$ where $B_{x}^{Q,0}$ and $B_{x}^{Q,1}$ are the amplitudes of the oscillating magnetic field, with $\omega_{Q}$ and $\phi$ being the frequency and phase, respectively. In experiments, the transverse oscillating field can be introduced by using the electron spin resonance \cite{Veldhorst.14,Veldhorst.15,Huang.19} or electron dipole spin resonance  \cite{Zajac.18, Zajac.18b,Yoneda.18,Watson.18} techniques. In the rotating frame and under the rotating-wave approximation, when  $\omega_{Q}$ matches the Larmor frequency, the Hamiltonian for each SSQ can be written as
\begin{equation}
\begin{aligned}
H_1(t)=&\frac{\Omega(t)}{2}(\rm{cos}\phi\ \sigma_{x}+\rm{sin}\phi\ \sigma_{y}),
\end{aligned}
\label{Hs}
\end{equation}
where $\Omega$ is the Rabi frequency related to the amplitude of the oscillating magnetic field. In experiments, both $\Omega$ and $\phi$ are time-dependent and can be controlled conveniently.

To implement the nonadiabatic single-qubit geometric gates, the entire evolution time is divided into three parts. In each part, the Rabi frequency $\Omega$ and the phase $\phi$ satisfy
\begin{equation}
\begin{aligned}
\int_{0}^{T_1}\Omega(\tau)d\tau&= \theta, \quad \left\{\phi_1=\phi-\frac{\pi}{2},\tau\in\left[0,T_1\right]\right\}\\
\int_{T_1}^{T_2}\Omega(\tau)d\tau&= \pi, \quad \left\{\phi_2=\phi+\gamma+\frac{\pi}{2},\tau\in\left[T_1,T_2\right]\right\}\\
\int_{T_2}^{T}\Omega(\tau)d\tau&= \pi-\theta, \quad \left\{\phi_3=\phi-\frac{\pi}{2},\tau\in\left[T_2,T\right]\right\},\\
\end{aligned}
\label{eq:seg}
\end{equation}
which leads to the evolution operator at the final time as
\begin{equation}
\begin{aligned}
U(\gamma,\theta,\phi)&=U(T,T_2)U(T_2,T_1)U(T_1,0)\\
%&=\cos\gamma+i\sin\gamma \left(\begin{array}{cc}\cos\theta&\sin\theta e^{-i \phi}\\ \sin\theta e^{i \phi}&-\cos\theta\end{array}\right)\\
&=e^{i\gamma\vec{n}\cdot \vec{\sigma}},
\end{aligned}
\label{eq:U_gs}
\end{equation}
where $\vec{n}=\left(\sin\theta\cos\phi,\sin\theta\sin\phi,\cos\theta\right)$ is the unit vector on the Bloch sphere with $0\leqslant \theta\leqslant \pi $ and $0\leqslant \phi<2\pi$. $\vec{\sigma}=\left(\sigma_{x},\sigma_{y},\sigma_{z}\right)$ are the Pauli matrixes.
% in the computational bases $\left|0\rangle\right.$ and $\left|1\rangle\right.$.
Thus, $U(\gamma,\theta,\phi)$ corresponds to  rotations around the axis $\vec{n}$ by an angle $-2\gamma$. Since all parameters here can be controlled independently, one is able to achieve arbitrary single-qubit operation.

It is straightforward to demonstrate that $U(\gamma,\theta,\phi)$ is the desired geometric gate by taking the two orthogonal states
\begin{equation}
\begin{aligned}
\left|\psi_{+}(t)\right\rangle&=\cos \frac{\theta(t)}{2}|0\rangle+\sin \frac{\theta(t)}{2} e^{i \phi(t)}|1\rangle,\\
\left|\psi_{-}(t)\right\rangle&=\sin \frac{\theta(t)}{2} e^{-i \phi(t)}|0\rangle-\cos \frac{\theta(t)}{2}|1\rangle,
\end{aligned}
\label{eq:Psipm}
\end{equation}
as the evolution states of the geometric gate.
% which are also the eigenstates of $\bm{n}\cdot\bm{\sigma}$.
After the nonadiabatic cyclic evolution (at final time $T$), these two orthogonal states acquire an extra global phase:
$U(\gamma,\theta,\phi)\left|\psi_{\pm}\right\rangle=e^{\pm  {i}\gamma} \left|\psi_{\pm}\right\rangle$. The evolution operator can then be rewritten as $U(\gamma,\theta,\phi)=e^{ {i}\gamma}\left|\psi_{+}\right\rangle\left\langle\psi_{+}\left| +e^{- {i} \gamma}\right| \psi_{-}\right\rangle\left\langle\psi_{-}\right|$, where the global phase $\gamma$ is determined by the solid angle enclosed by the cyclic evolution \cite{Erik.15}.
Here, we elaborate on how the geometric phase can be achieved by the evolution state $\left|\psi_{+}(t)\right\rangle$, and that for $\left|\psi_{-}(t)\right\rangle$ can be understood in a similar way. As seen in Fig.~\ref{fig:qdot_bloch}(b), in the first part of the evolution, $\left|\psi_{+}(0)\right\rangle$ starts from a given point A on the Bloch sphere, corresponding to $\theta=\theta(0)$ and $\phi=\phi(0)$, and evolves along the geodesic line up to the north pole B. In the second part, it goes down to the south pole D along another geodesic line, which is $\gamma$ apart from the previous one. Finally, it goes back to the starting point at the end of the third part. It is shown that after the cyclic evolution, the state $\left|\psi_{+}(t)\right\rangle$ traces out an orange-slice-shaped loop $A-B-C-D-A$. Since the evolution is always along geodesic lines, the dynamical phase is cancelled out (see Appendix \ref{appx:phase}).
%Thus, $\gamma$ is a pure geometric phase: \underline{$\gamma=-\frac{\Gamma }{2}$} where $\Gamma $ is the solid angle enclosed by the loop.
Besides, we can see that the parallel-transport condition also satisfies $\left\langle\psi_{\pm}\left|U^{\dagger}(\gamma,\theta,\phi) H_{1} U(\gamma,\theta,\phi)\right| \psi_{\pm}\right\rangle= 0$ \cite{Zhao.17}. Therefore, we conclude that $U(\gamma,\theta,\phi)$ represents a pure geometric gate.

\subsection{Geometric two-qubit gate}
Next, we show how to realize the geometric two-qubit gate. As depicted in Fig.~\ref{fig:qdot_bloch}(a), the neighboring two spins confined in the two quantum dots are coupled by the exchange interaction $J$. The corresponding  Hamiltonian is \cite{Zajac.18b}
\begin{equation}
H_2(t)=J(t)(\bm{S}_{L}\cdot\bm{S}_{R}-1/4)+\bm{S}_{L}\cdot\bm{B}_{L}+\bm{S}_{R}\cdot\bm{B}_{R},
\label{Ht}
\end{equation}
Here, ${S}_{L}$ and ${S}_{R}$ denote the spin of the electron in the left and right quantum dot, respectively. In the two-qubit basis $\{\left|0 0\rangle\right.,\left|0 1\rangle\right.,\left|1 0\rangle\right.,\left|1 1\rangle\right.\}$
%($\{\left|\uparrow \uparrow\rangle\right.,\left|\uparrow \downarrow\rangle\right.,\left|\downarrow \uparrow\rangle\right.,\left|\downarrow \downarrow\rangle\right.\}$)
the Hamiltonian in Eq.~(\ref{Ht}) is
\begin{equation}
\begin{aligned}
H_2(t)=&\left(\begin{array}{cccc}
{E_{z}+J(t)/2} & {0} & {0} & {0} \\
{0} & {\delta E_{z} / 2} & {J(t) / 2} & {0} \\
{0} & {J(t) / 2} & -\delta E_{z} / 2 & {0} \\
{0} & {0} & {0} & {-E_{z}+J(t)/2}\end{array}\right),
\end{aligned}
\label{Htmatrix}
\end{equation}
where $E_{z}=B_{z}^{h}+(B_{z}^{L}+B_{z}^{R})/2$ and $\delta E_{z}=B_{z}^{L}-B_{z}^{R}$. Note that we have lifted the zero-point energy by $J/2$ and shut down the oscillating magnetic field.
%Experimentally, the exchange interaction $J$ can be conveniently controlled by gate voltages and we always have $J \ll E_{z}, \delta E_{z}$ \cite{Zajac.18, Zajac.18b} when implementing the two-qubit operation. So that, $J(t)$ can be treated as a perturbation.
Now, $H_2(t)$ can be further divided into two parts $H_2(t)=H^{0}_{2}+H_2'$ where $H_2'=\frac{J(t)}{2} \left|\uparrow \downarrow\rangle\right.\left \langle  \downarrow \uparrow \right| + h.c.$ is the perturbation term and $H^{0}_{2}$ is the remaining free Hamiltonian. Therefore, in the interaction picture $H_{2}(t)$ can be rewritten as
\begin{equation}\label{HtImatrix}
H_{2I}(t)=\frac{J(t)}{2} \left( e^{i\delta E_{z}t} |0 1\rangle \langle 1 0|+h.c.\right).
\end{equation}
Further, we assume that the exchange interaction $J(t)$ is operated in an oscillating way \cite{sigillito.19,Kenta.19,Nichol.17} as $J(t)=j_0+j(t)\rm{cos}(\omega_{j}t+\psi)$. Since both $j_0, j(t) \ll \delta E_z$, in the rotating-wave approximation, Eq.~(\ref{HtImatrix}) reduces to
\begin{equation}
\begin{aligned}
H_{2R}(t)=&\frac{j(t)}{2}(\rm{cos}\psi\tilde{\sigma}_{x}+\rm{sin}\psi\tilde{\sigma}_{y}).
\end{aligned}
\label{H2t}
\end{equation}
where $\tilde{\sigma}_{x}$ and $\tilde{\sigma}_{y}$ are the effective Pauli matrixes in the $\{\left|0 1\rangle\right.,\left|1 0\rangle\right.\}$ subspace. Similar to the single-qubit case, we can construct the two-qubit geometric gate by dividing the evolution time into three segments.
\begin{equation}
\begin{aligned}
\int_{0}^{T_1}j(\tau)d\tau&= \vartheta, \quad \left\{\psi_1=\psi-\frac{\pi}{2},\tau\in\left[0,T_1\right]\right\}\\
\int_{T_1}^{T_2}j(\tau)d\tau&= \pi, \quad \left\{\psi_2=\psi+\xi+\frac{\pi}{2},\tau\in\left[T_1,T_2\right]\right\}\\
\int_{T_2}^{T}j(\tau)d\tau&= \pi-\vartheta, \quad \left\{\psi_3=\psi-\frac{\pi}{2},\tau\in\left[T_2,T\right]\right\},\\
\end{aligned}
\label{eq:seg2}
\end{equation}
In this way, the achieved evolution operator is
\begin{equation}
\begin{aligned}
&U_2(\xi,\vartheta,\psi)=U_{2}(T,T_2)U_{2}(T_2,T_1)U_{2}(T_1,0)\\
&=
\left(\begin{array}{cccc}
{1}  & {0} & {0} & {0}  \\
 {0} & {\cos \xi+i \sin \xi \cos \vartheta} & {i \sin \xi \sin \vartheta e^{-i \psi}} & {0} \\
{0} & {i \sin \xi \sin \vartheta e^{i \psi}} & {\cos \xi-i \sin \xi \cos \vartheta} & {0} \\
 {0} & {0} & {0} & {1}\end{array}\right),
\end{aligned}
\label{eq:U_gt}
\end{equation}
with $0\leqslant \vartheta\leqslant \pi $ and $0\leqslant \psi < 2\pi $. It is easy to find that by setting $\vartheta=\xi=\pi/2$ and $\psi=0$, Eq.~(\ref{eq:U_gt}) is equivalent to an $\rm{iSWAP}$ gate.
%\begin{equation}
%\begin{aligned}
%\rm{iSWAP}&=U_2(\pi/2,\pi/2,0)=\left(\begin{array}{cccc}
%{1}  & {0} & {0} & {0}  \\
%{0} & 0 & i & {0} \\
%{0} & i  & 0 & {0} \\
%{0} & {0} & {0} & {1}
%\end{array}\right).
%\end{aligned}
%\label{eq:iSWAP}
%\end{equation}
On the other hand, since $H_{2R}(t)$ actually belongs to the XY-interacted Hamiltonian \cite{Schuch.03} where the Hamiltonian only appears in the effective $\sigma_{x}$ and $\sigma_{y}$ terms in the considered subspace. For comparison, one can also construct the iSWAP gate in the dynamical way by only one step $\mathrm{iSWAP}\equiv\exp[-i H_{2R} \frac{\pi}{j}]$. In fact, the iSWAP gate can be regarded as half of the CNOT gate \cite{Schuch.03},
%. As seen in Fig.~\ref{fig:cnot},
i.e., the CNOT gate can be obtained by applying the iSWAP gate twice combined with appropriate single-qubit operations.

%\begin{figure}
%	\includegraphics[width=1.03\columnwidth]{Fig2.pdf}
%	\caption{CNOT gate obtained by implementing twice of the iSWAP gate combined with several single-qubit gates.}
%	\label{fig:cnot}
%\end{figure}

\section{Robustness proof}\label{sec:comparison}

Here, we demonstrate how geometric gates are less sensitive to the considered noises compared to their dynamical counterparts. Before turning to this demonstration, we show how to obtain arbitrary SSQ dynamical rotations. In the absence of noise, the dynamical operator corresponding to the Hamiltonian in Eq.~(\ref{Hs}) is
\begin{equation}
\begin{aligned}
R(\bm{r}, \gamma)&\equiv \exp\left[-i\frac{\gamma}{2}(\rm{cos}\phi\ \sigma_{x}+\rm{sin}\phi\ \sigma_{y})\right],
\end{aligned}
\label{eq:Rot}
\end{equation}
where $R(\bm{r}, \gamma)$ denotes the rotation in the $x$-$y$ plane by an angle $\gamma$, and the rotation axis $\bm{r}$ in the plane is determined by $\phi$. As the Rabi frequency $\Omega$ is assumed to be constant, $R(\bm{r}, \gamma)$ is actually a single-piece rotation. Arbitrary single-qubit gate requires two nonparallel axis rotation on the Bloch sphere. This can be implemented by setting $\phi$ to be 0 ($\pi$) and $\frac{\pi}{2}$ ($-\frac{\pi}{2}$), which corresponds to rotations around $\hat{x}$ (-$\hat{x}$) and $\hat{y}$ ($-\hat{y}$) axis, respectively. For other rotations out of the plane, one can decompose it into a $\hat{x}$-$\hat{y}$-$\hat{x}$ composite pulse sequence
\begin{equation}
\begin{aligned}
R(\hat{x},\gamma_a)R(\hat{y},\gamma_b)R(\hat{x},\gamma_c),
\end{aligned}
\label{eq:Rdecx}
\end{equation}
or a $\hat{y}$-$\hat{x}$-$\hat{y}$ sequence
\begin{equation}
\begin{aligned}
R(\hat{y},\gamma_a)R(\hat{x},\gamma_b)R(\hat{y},\gamma_c).
\end{aligned}
\label{eq:Rdecy}
\end{equation}
One should not confuse the composite pulses here with the geometric gates, which is also composed of multiple pieces.  First, for the geometric gate, the Rabi frequency $\Omega$ and the phase $\phi$ are divided into three distinct parts so as to form the cyclic evolution loop, while it is not required to do so for the dynamical gate. Second, each part in the geometric gate has strong intrinsic connection to ensure that the dynamical phase can be cancelled during the evolution, and only the desired geometric phase remains. In this way we can use the merit of the geometric phase, which can be demonstrated less sensitive to the noise as shown later. In contrast, the dynamical rotation directly uses the noncyclic dynamical phases, thus, is more vulnerable to noise that are concerned in silicon.

Next, we compare the fidelity between the geometric gates and the dynamical counterparts in the noise environments. We assume that a SSQ is suffering from two types of noises, namely, the detuning noise and the systematic noise. Thus, the dynamical rotation subjected to noise reads
\begin{equation}
R(\bm{r}, \gamma)
= \exp\left[-i\left\{(1+\epsilon)H_1(t)+\delta\sigma_{z}\right\}\frac{\gamma}{\Omega}\right],
\label{eq:Rotn}
\end{equation}
where $\delta$ and $\epsilon$ are the detuning  and systematic noises, respectively. In experiments, systematic noise leads to the error in the Rabi oscillation while the detuning noise results in the lift or decline of the energy level, namely, detuning, which is assumed to be the dominant noise for silicon-based spin qubits.

For simplicity, we assume that both the systematic noise and the detuning noise are constant during the gate operation, and are independent from each other. We also assume the noises are weak enough compared to the Rabi oscillation, $\left |\delta  \right | \ll\frac{1}{T}\int_{0}^{T}\Omega(t)dt$  and $\left |\epsilon  \right | \ll\frac{1}{T}\int_{0}^{T}\Omega(t)dt$. Then, the fidelity of the $\hat{x}$-axis rotation according to Eq.~(\ref{eq:U_gs}) and (\ref{eq:Rotn}) can be expanded as
\begin{equation}
\begin{aligned}
\mathcal{F}_{1,d}(\hat{x},\gamma)&= 1-\frac{\gamma^{2}}{8}\epsilon^{2}+(\rm{cos}\gamma-1)\delta^{2},\\ %+O(\epsilon^{3}+\delta^{3})
\mathcal{F}_{1,g}(\hat{x},\gamma)&= 1-\frac{\pi^{2}\rm{sin}^{4}(\gamma/4)}{2}\epsilon^{2}-8\rm{cos}^{4}(\delta/4)\delta^{2},
%\\ &+O(\epsilon^{3}+\delta^{3}),
\end{aligned}
\label{eq:expand}
\end{equation}
with respect to $\epsilon$ and $\delta$ up to second order. Here, $\mathcal{F}_{1,d}(\hat{x},\gamma)$ and $\mathcal{F}_{1,g}(\hat{x},\gamma)$ denote the fidelity for the dynamical and geometric gate, respectively. Then, the fidelity difference between them is
\begin{equation}
\begin{aligned}
\Delta \mathcal{F}_1&=F_{1,g}(\hat{x},\gamma)-F_{1,d}(\hat{x},\gamma)=\Delta \mathcal{F}_{1,\epsilon}+\Delta \mathcal{F}_{1,\delta},\\
\Delta \mathcal{F}_{1,\epsilon}&=\frac{1}{8}[\gamma^{2}-4\pi^{2}\rm{sin}^{4}(\gamma/4)]\epsilon^{2},\\
\Delta \mathcal{F}_{1,\delta}&=-2[1+2\rm{cos}(\gamma/2)+\rm{\cos}\gamma]\delta^{2},
\end{aligned}
\label{eq:gd}
\end{equation}
where $\Delta \mathcal{F}_{1,\epsilon}$ and $\Delta \mathcal{F}_{1,\delta}$ denote the case with respect to the systematic noise and detuning noise, respectively. One can easily verify that, $\Delta \mathcal{F}_{1,\epsilon}\geqslant0$ in the regime $-\pi\leqslant \gamma\leqslant \pi $ which is enough to perform the desired rotation angle. This means the geometric gate is less sensitive to the systematic noise compared to the dynamical one. Unfortunately, $\Delta \mathcal{F}_{1,\delta}\leqslant0$ in this region, suggesting that the geometric gate performs worse than the dynamical one when the detuning noise is dominant. This fact %that the geometric gate can improve the systematic noise (appear in the diagonal term of the Hamiltonian) but incapable of action for the detuning noise (appear in the offdiagonal term of the Hamiltonian)
implies that the performance of geometric gate may be affected by the detailed noise structure in the Hamiltonian.

\begin{figure}
	\includegraphics[width=1.0\columnwidth]{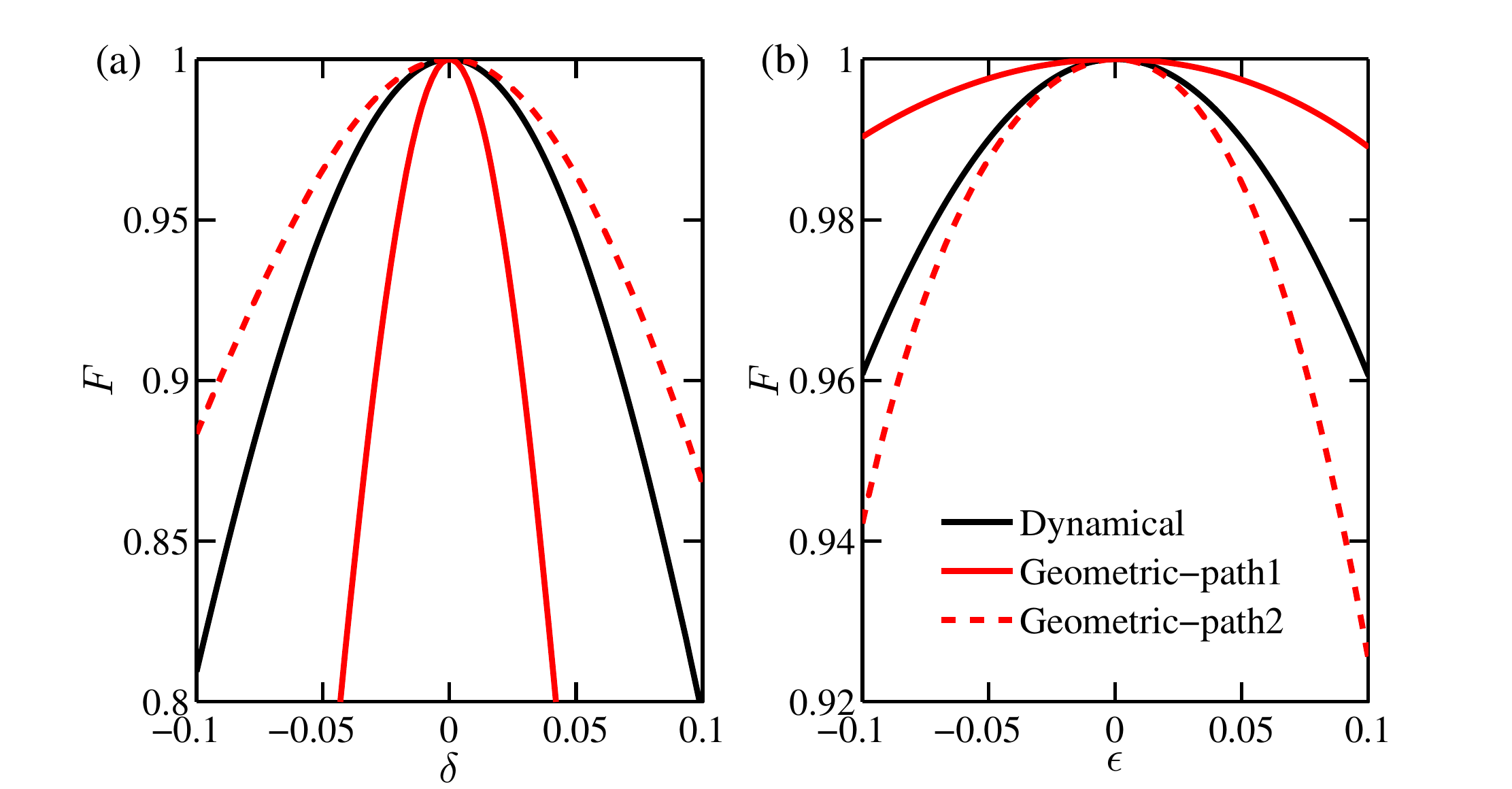}
	\caption{Fidelity of the CNOT gate as a function of the systematic noise in panel (a) and detuning noise in panel (b). The black solid lines denote the results for the dynamical gates, while the solid and dashed red lines indicate the geometric gate in path 1 and path 2, respectively.}
	\label{fig:Fcnot}
\end{figure}

Therefore, one possible way is to seek for another evolution path to acquire the geometric phase, which may be immune to the detuning noise. Observing the evolution path on the Bloch sphere as shown in Fig.~\ref{fig:qdot_bloch}(b), by modulating $\phi_2$ in Eq.~(\ref{eq:seg}), we are allowed to select various geodesic lines to evolve the orthogonal state during the second part of evolution. Here, we take the evolution of $\left|\psi_{+}(t)\right\rangle$ to explain. By choosing $\phi_{2}'=\phi_{2}-\gamma'$,  $\left|\psi_{+}(t)\right\rangle$ can go along a new path $B-{C}'-D$, and thus traces out another enclosed loop $A-B-{C}'-D-A$. The geometric phase then turns from $\gamma$ to $\gamma-\gamma'$, and the corresponding new geometric operator becomes
\begin{equation}
\begin{aligned}
U'(\gamma',\gamma,\theta,\phi)&=e^{i(\gamma-\gamma')\bm{n}.\bm{\sigma}}.
\end{aligned}
\label{eq:U_gs'}
\end{equation}
As a simple but profound example, we take $\gamma'=\pi$, while keeping other parameters unchanged. As shown in Fig.~\ref{fig:qdot_bloch}(b), we denote the new path as ``path 2'' (in dashed-red line) and the original one as ``path 1'' (in solid-blue line) for comparison. One can see that the geodesic line related to path 2 is tilted by $\pi$ from the original one. In this case, we have $\gamma_{2}=\gamma_{1}-\pi$, and the corresponding solid angle is changed from $-2\gamma$ to $-2\gamma+2\pi$. So that, the target evolution operator $U'(\pi,\gamma,\theta,\phi)$
is equivalent to $U(\gamma,\theta,\phi)$ (except a negative sign). We surprisingly find that, although the new operator remains the same as before, the noise-resilience for them can be substantially different. This can be seen by further expanding the fidelity of the new operator in path 2 and we have
\begin{equation}
\begin{aligned}
%\Delta \mathcal{F}_{2}^{2}&=\mathcal{F}_{2,g}^{2}(\hat{x},\gamma)-\mathcal{F}_{2,d}^{2}(\hat{x},\gamma)\\
\Delta \mathcal{F}_{2,\epsilon}^{2}&=\frac{1}{4}(\gamma^{2}-4\pi^{2}\rm{cos}^{4}(\gamma/4))\epsilon^{2},\\
\Delta \mathcal{F}_{2,\delta}^{2}&=(8\rm{cos}(\gamma/2)-4(1+\rm{\cos}\gamma))\delta^{2}.\\
\end{aligned}
\label{eq:gd2}
\end{equation}
Here, for the considered region $-\pi\leqslant \gamma\leqslant \pi $, we always have $\Delta \mathcal{F}_{2,\epsilon}^{2}\leqslant0$ and $\Delta \mathcal{F}_{2,\delta}^{2}\geqslant 0$. Therefore, the new operator related to path 2 is robust against the detuning noise but is more sensitive to the systematic noise. This result is inverse to the previous geometric gate in path 1. Thus, if the systematic noise is dominant, we may select path 1 to construct the geometric gate, otherwise take path 2 when the qubit suffers mainly from the detuning noise. Note that, the expansion results for the $\hat{y}$-axis rotation is similar and we are not going to show the detail again. On the other hand, since the Hamiltonian related to the two-qubit gate acts the way like the single-qubit case in the $\{\left|0 1\rangle\right.,\left|1 0\rangle\right.\}$ subspace, we can also analyze it using the same method. Therefore, arbitrary single- and two-qubit dynamical gates can be improved by the proper geometric gates. For the construction of the CNOT gate, both the single-qubit gates and the two-qubit iSWAP gate are involved. Thus, we plot the fidelity of the CNOT gate to show the advantage of the geometric gates as shown in Fig.~\ref{fig:Fcnot}. It is clearly shown there, for both the detuning noise and the systematic noise, the fidelity of the dynamical CNOT gate can be improved by the proper geometric gates.

\begin{table*}[h]
	
	\caption{Clifford gates used in the randomized benchmarking simulation}	
	\label{tab:rb}
	%\footnotesize
	\centering
	\begin{tabular}{lll}
		\hline\noalign{\smallskip} Clifford element & Dynamical & Geometric  \\
		\noalign{\smallskip}\hline\noalign{\smallskip}		
		$C_0=\hat{I}$ & $R(\hat{x},2\pi)$  &  $U(-\pi, \frac{\pi}{2}, 0)$\\
		
		$C_1=R(\hat{x},-\frac{\pi}{2})$ & $R(-\hat{x},\frac{\pi}{2})$ &  $U(-\frac{\pi}{4}, \frac{\pi}{2}, \pi)$  \\
		
		$C_2=R(\hat{x},\frac{\pi}{2})$ & $R(\hat{x},\frac{\pi}{2})$ &  $U(-\frac{\pi}{4}, \frac{\pi}{2}, 0)$  \\
		
		$C_3=R(\hat{x},\pi)$ & $R(\hat{x},\pi)$  &  $U(-\frac{\pi}{2}, \frac{\pi}{2}, 0)$  \\
		
		$C_4=R(\hat{y},-\frac{\pi}{2})$ & $R(-\hat{y},\frac{\pi}{2})$  &  $U(-\frac{\pi}{4}, \frac{\pi}{2}, -\frac{\pi}{2})$   \\
		
		$C_5=R(\hat{y},\frac{\pi}{2})$ & $R(\hat{y},\frac{\pi}{2})$  &  $U(-\frac{\pi}{4}, \frac{\pi}{2}, \frac{\pi}{2})$  \\
		
		$C_6=R(\hat{y},\pi)$ & $R(\hat{y},\pi)$  &  $U(-\frac{\pi}{2}, \frac{\pi}{2}, \frac{\pi}{2})$   \\
		
		$C_7=R(\hat{z},-\frac{\pi}{2})$ & $R(\hat{x},\frac{\pi}{2})$ $R(-\hat{y},\frac{\pi}{2}) R(-\hat{x},\frac{\pi}{2})$ & $U(-\frac{\pi}{4}, \frac{\pi}{2}, 0) U(-\frac{\pi}{4}, \frac{\pi}{2}, -\frac{\pi}{2}) U(-\frac{\pi}{4}, \frac{\pi}{2}, \pi)$  \\
		
		$C_8=R(\hat{z},\frac{\pi}{2})$ & $R(\hat{x},\frac{\pi}{2})$ $R(\hat{y},\frac{\pi}{2}) R(-\hat{x},\frac{\pi}{2})$ &  $U(-\frac{\pi}{4}, \frac{\pi}{2}, 0) U(-\frac{\pi}{4}, \frac{\pi}{2}, \frac{\pi}{2}) U(-\frac{\pi}{4}, \frac{\pi}{2}, \pi)$  \\
		
		$C_9=R(\hat{z},\pi)$ & $R(\hat{x},\pi) R(\hat{y},\pi) $  &  $U(-\frac{\pi}{2}, \frac{\pi}{2}, 0) U(-\frac{\pi}{2}, \frac{\pi}{2}, \frac{\pi}{2}) $  \\
		
		$C_{10}=R(\hat{x}+\hat{z},\pi)$ & $R(-\hat{y},\frac{\pi}{2}) R(\hat{x},\pi)$ & $U(-\frac{\pi}{4}, \frac{\pi}{2}, -\frac{\pi}{2}) U(-\frac{\pi}{2}, \frac{\pi}{2}, 0) $ \\
		
		$C_{11}=R(\hat{x}-\hat{z},\pi)$ & $R(\hat{y},\frac{\pi}{2}) R(\hat{x},\pi)$ & $U(-\frac{\pi}{4}, \frac{\pi}{2}, \frac{\pi}{2}) U(-\frac{\pi}{2}, \frac{\pi}{2}, 0) $  \\
		
		$C_{12}=R(\hat{x}+\hat{y},\pi)$ & $R(\hat{x},\frac{\pi}{2})$ $R(\hat{y},\frac{\pi}{2}) R(\hat{x},\frac{\pi}{2})$ & $U(-\frac{\pi}{4}, \frac{\pi}{2}, 0) U(-\frac{\pi}{4}, \frac{\pi}{2}, \frac{\pi}{2}) U(-\frac{\pi}{4}, \frac{\pi}{2}, 0)$ \\
		
		$C_{13}=R(\hat{x}-\hat{y},\pi)$ & $R(\hat{x},\frac{\pi}{2})$ $R(-\hat{y},\frac{\pi}{2}) R(\hat{x},\frac{\pi}{2})$ & $U(-\frac{\pi}{4}, \frac{\pi}{2}, 0) U(-\frac{\pi}{4}, \frac{\pi}{2}, -\frac{\pi}{2}) U(-\frac{\pi}{4}, \frac{\pi}{2}, 0)$  \\
		
		$C_{14}=R(\hat{y}+\hat{z},\pi)$ & $R(\hat{x},\frac{\pi}{2}) R(\hat{y},\pi)$ & $U(-\frac{\pi}{4}, \frac{\pi}{2}, 0) U(-\frac{\pi}{2}, \frac{\pi}{2}, \frac{\pi}{2}) $  \\
		
		$C_{15}=R(\hat{y}-\hat{z},\pi)$ & $R(-\hat{x},\frac{\pi}{2}) R(\hat{y},\pi)$ & $U(-\frac{\pi}{4}, \frac{\pi}{2}, \pi) U(-\frac{\pi}{2}, \frac{\pi}{2}, \frac{\pi}{2}) $  \\
		
		$C_{16}=R(\hat{x}+\hat{y}+\hat{z},\frac{2\pi}{3})$ & $R(\hat{x},\frac{\pi}{2}) R(\hat{y},\frac{\pi}{2})$  & $U(-\frac{\pi}{4}, \frac{\pi}{2}, 0) U(-\frac{\pi}{4}, \frac{\pi}{2}, \frac{\pi}{2}) $  \\
		
		$C_{17}=R(\hat{x}+\hat{y}+\hat{z},\frac{4\pi}{3}$ & $R(-\hat{y},\frac{\pi}{2}) R(-\hat{x},\frac{\pi}{2})$& $U(-\frac{\pi}{4}, \frac{\pi}{2}, -\frac{\pi}{2}) U(-\frac{\pi}{4}, \frac{\pi}{2}, \pi) $  \\
		
		$C_{18}=R(\hat{x}+\hat{y}-\hat{z},\frac{2\pi}{3})$ & $R(\hat{y},\frac{\pi}{2}) R(\hat{x},\frac{\pi}{2})$ & $U(-\frac{\pi}{4}, \frac{\pi}{2}, \frac{\pi}{2}) U(-\frac{\pi}{4}, \frac{\pi}{2}, 0) $  \\
		
		$C_{19}=R(\hat{x}+\hat{y}-\hat{z},\frac{4\pi}{3})$ & $R(-\hat{x},\frac{\pi}{2}) R(-\hat{y},\frac{\pi}{2})$ & $U(-\frac{\pi}{4}, \frac{\pi}{2}, \pi) U(-\frac{\pi}{4}, \frac{\pi}{2}, -\frac{\pi}{2}) $  \\
		
		$C_{20}=R(\hat{x}-\hat{y}+\hat{z},\frac{2\pi}{3})$ & $R(-\hat{y},\frac{\pi}{2}) R(\hat{x},\frac{\pi}{2})$  & $U(-\frac{\pi}{4}, \frac{\pi}{2}, -\frac{\pi}{2}) U(-\frac{\pi}{4}, \frac{\pi}{2}, 0) $  \\
		
		$C_{21}=R(\hat{x}-\hat{y}+\hat{z},\frac{4\pi}{3})$ & $R(-\hat{x},\frac{\pi}{2}) R(\hat{y},\frac{\pi}{2})$& $U(-\frac{\pi}{4}, \frac{\pi}{2}, \pi) U(-\frac{\pi}{4}, \frac{\pi}{2}, \frac{\pi}{2}) $  \\
		
		$C_{22}=R(-\hat{x}+\hat{y}+\hat{z},\frac{2\pi}{3})$ & $R(\hat{y},\frac{\pi}{2}) R(-\hat{x},\frac{\pi}{2})$& $U(-\frac{\pi}{4}, \frac{\pi}{2}, \frac{\pi}{2}) U(-\frac{\pi}{4}, \frac{\pi}{2}, \pi) $   \\
		
		$C_{23}=R(-\hat{x}+\hat{y}+\hat{z},\frac{4\pi}{3})$ & $R(\hat{x},\frac{\pi}{2}) R(-\hat{y},\frac{\pi}{2})$ & $U(-\frac{\pi}{4}, \frac{\pi}{2}, 0) U(-\frac{\pi}{4}, \frac{\pi}{2}, -\frac{\pi}{2}) $   \\		
		\noalign{\smallskip}\hline
		
	\end{tabular}
	
\end{table*}

\begin{figure}
	\includegraphics[width=0.66\columnwidth]{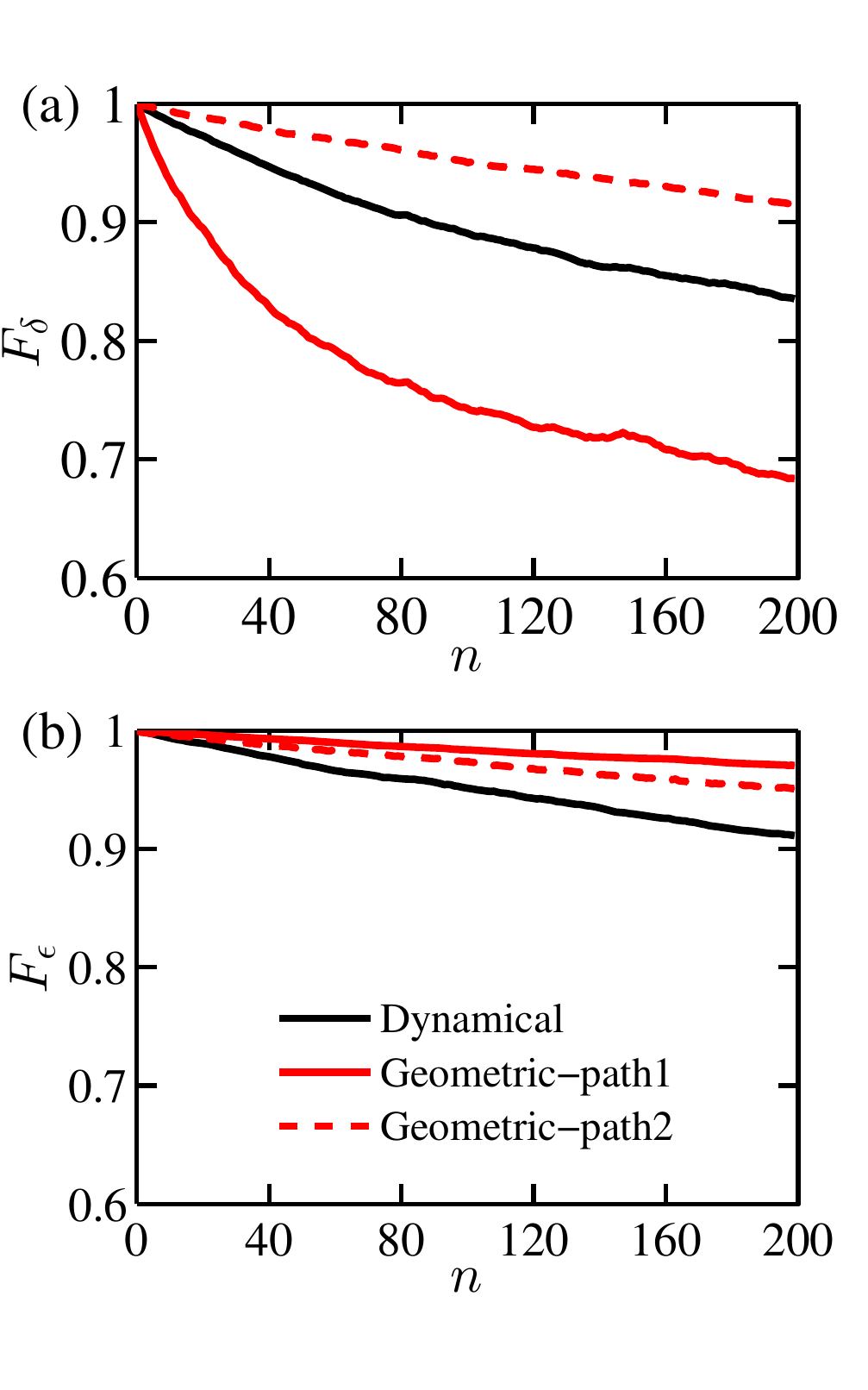}
	\caption{Results of the RB for the single-qubit Clifford gates under the static noise model. The results in panel (a) and (b) are with respect to the detuning noise and systematic noise, respectively.}
	\label{fig:rbstatic}
\end{figure}

\section{numerical verification of robustness}\label{sec:rb}

To test the superiority of the geometric gates, we carry out RB \cite{Magesan.12, Knill.08, Zhang.16} which is a powerful technique to investigate gate fidelities under specific noise conditions. In RB, instead of studying arbitrary single-qubit gate, we  focus ourselves on a finite subset, i.e. the Clifford group \cite{Mark.02}. The RB process is implemented by averaging the gate fidelity over gate sequences randomly drawn from the Clifford group and over different noise realizations. Then, we can quantitatively compare the performance of the dynamical gates and the geometric gates. In Table.~\ref{tab:rb}, we show how to construct the Clifford group for the two kinds of gates, both of which includes 24 elements.
For the dynamical Clifford elements, except for the $\hat{x}$- and $\hat{y}$-axis rotations, each Clifford element can be divided into the combination of $R(\hat{x},\gamma)$ and $R(\hat{y},\gamma)$. We find that the mean gate number per Clifford element is 1.875. To compare the dynamical gates and geometric gates fairly, each $R(\hat{x},\gamma)$ or $R(\hat{y},\gamma)$ in the dynamical element is replaced by the corresponding geometric gate.

We first consider the static noise model. In our RB simulation, $\epsilon$ and $\delta$ in each run are drawn from the Gaussian distribution, namely, $\sigma_{\epsilon}^{2}: \mathcal{N}\left(0, \sigma_{\epsilon}^{2}\right)$ and $\sigma_{\delta}^{2}: \mathcal{N}\left(0, \sigma_{\delta}^{2}\right)$, where $\sigma_{\epsilon}^{2}$ and $\sigma_{\delta}^{2}$ are the variance with respect to the noise. In Ref.~\cite{Huang.19}, the typical Rabi frequency is about 500 kHz, and the variance of the detuning noise is 10$\sim$20 kHz. Such that, we consider $\sigma_{\delta}=\sigma_{\epsilon}=0.02$ for simulation. The averaged fidelity is obtained by fitting the resulting fidelity curve to $\left(1+e^{-d n}\right)/2$, where $d$ denotes the averaged error per gate, and $n$ is the Clifford gate number. In Fig.~\ref{fig:rbstatic}(a), we show the benchmarking results when only the detuning noise is present. We find that, the geometric gates in path 2 outperform the dynamical gates, where the averaged fidelity is 0.9990. However, those gates in path 1 perform even worse than the dynamical ones, since they are highly sensitive to the detuning noise. When only the systematic noise is present [Fig.~\ref{fig:rbstatic}(b)], the geometric gate in both paths perform better than the dynamical ones. We also find that the geometric gates in path 1 are standing out, and the fidelity can be as high as 0.9997.

\begin{figure}
	\includegraphics[width=1.0\columnwidth]{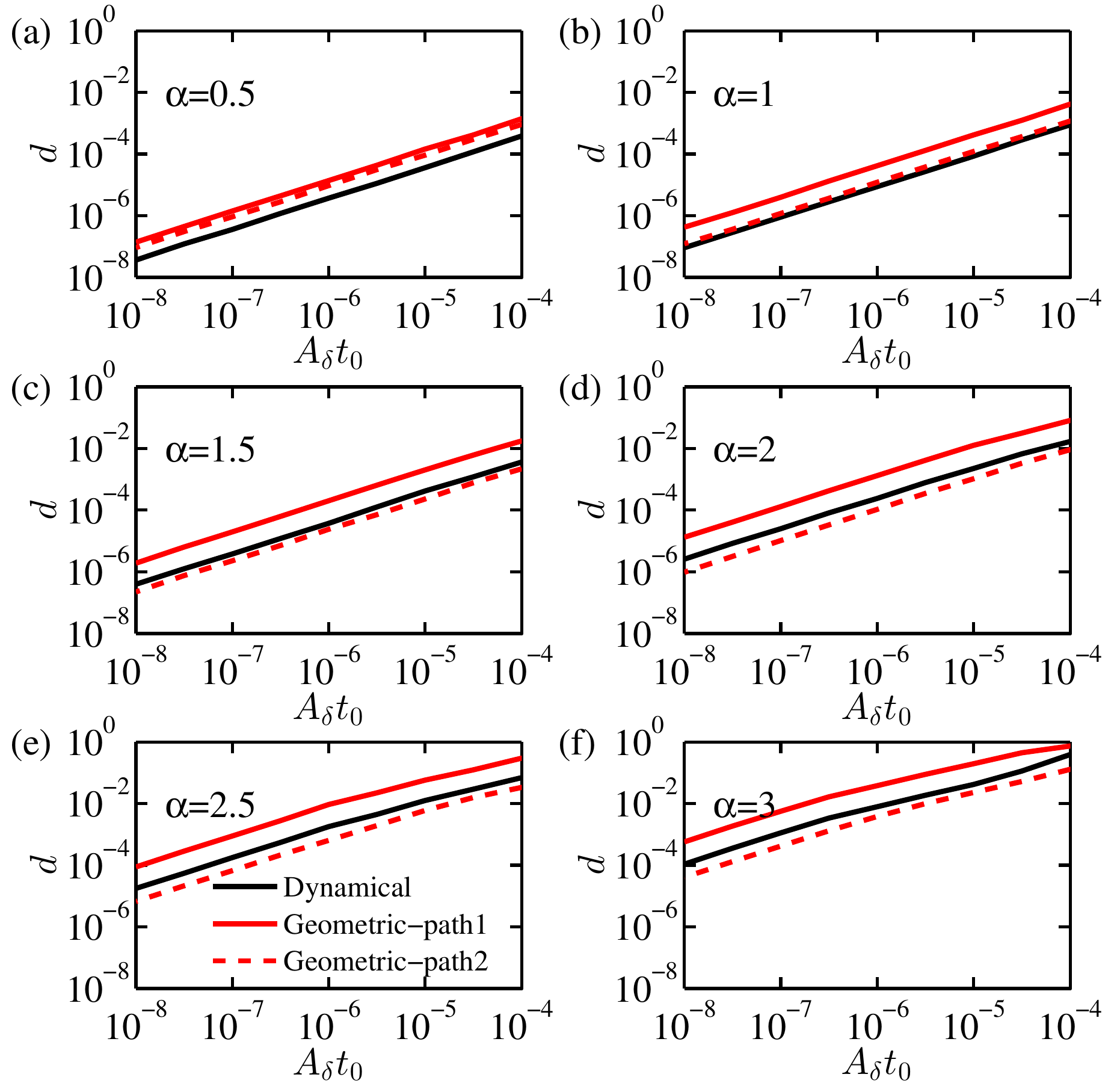}
	\caption{Average error per gate $d$ vs. noise amplitude $A_{\delta}t_0$ for $1/f$-type noise with different noise exponent $\alpha$. $d$ is obtained by fitting the fidelity curves resulting from the RB.}
	\label{fig:gammad}
\end{figure}

Although the geometric gates are superior to the dynamical gates in the static-noise model, whether it can persist for the time-dependent noise remains unknown. In experiments, the time-dependent noises are commonly modeled by the $1/f$ type, whose power spectral density has the form as $S(\omega)=A/(\omega t_0)^\alpha$. Here, $A$ represents the amplitude of the noise, the exponent $\alpha$ denotes how much the noise is correlated crucial to characterize the noise, and $t_0$ is an arbitrary time unit, depending on the magnitude of the Rabi frequency. In this work, we are using the method in Refs.~\cite{Yang.16,Zhang.16} to simulate the $1/f$ noise, where we are able to generate noise spectrum with $0 \leqslant \alpha \leqslant 3$. In Ref. \cite{Zhang.16}, we have shown that, for small $\alpha$, the noise is not correlated at all, and is closed to the white-like noise. While for the large $\alpha$, the randomness of the noise is reducing and the noise is rising or lowering in a relative long time scale, so that, the noise is closed to the quasi-static model. Therefore, we may expect the better performance of the geometric gates for the large $\alpha$.

\begin{figure}
	\includegraphics[width=1.0\columnwidth]{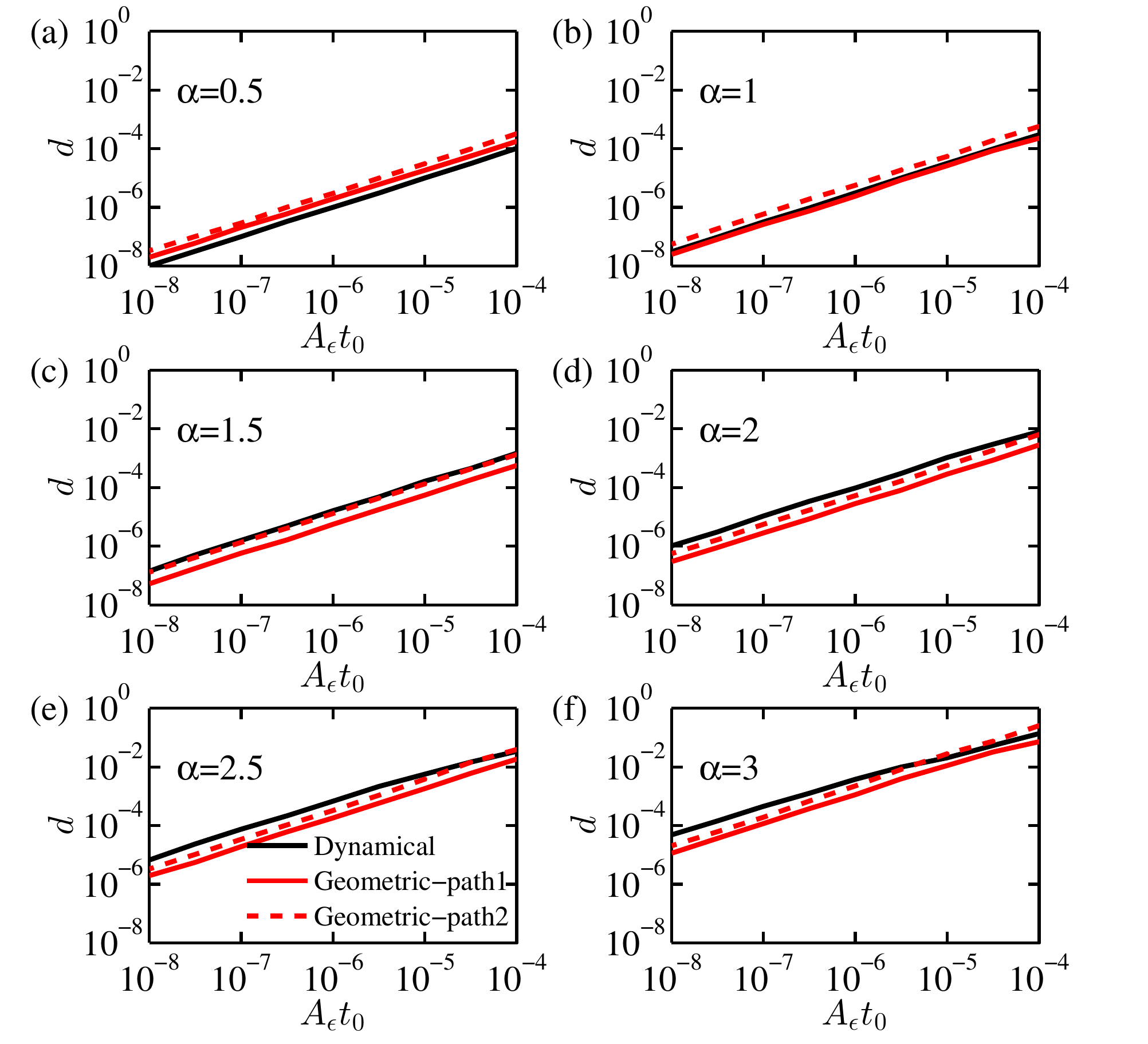}
	\caption{Average error per gate $d$ vs. noise amplitude $A_{\epsilon}t_0$ for $1/f$-type noise with different noise exponent $\alpha$. $d$ is obtained by fitting the fidelity curves resulting from the randomized benchmarking.}
	\label{fig:gammae}
\end{figure}

In Fig.~\ref{fig:gammad}, we show the dependence of the error $d$ on the detuning noise amplitude $A_\delta$ for both the dynamical and geometric gates. When $\alpha$ is small enough, i.e. $\alpha=0.5$, one can see that the error for the geometric gates are even larger than the dynamical ones. This means the geometric gates can not offer any improvement, because the noise is far away from the static model due to the small $\alpha$. When it comes to the intermediate value of $\alpha=1$, the geometric gates in path 2 start to offer improvement against the dynamical counterparts, and the improvement is becoming more and more pronounced as $\alpha$ keeps increasing. We also see that for any $\alpha$ value, the errors for the geometric gates in path 1 are the largest. This result is consistent with the case in the static noise model. According to the experimental data in Ref. \cite{Chan.18}, the $\alpha$ value of the detuning noise at the low-frequency domain is 2.5. And, the noise spectrum there has been measured to be $S(\omega)\approx C_1/\omega^{2.5}$ with $C_1=3\times 10^{13}$. Converting this data to our unit, we have $At_0=C_1 t_{0}^{\alpha+1} $. If we take the Rabi frequency to be 500 kHz, which means $t_0=2\mu s$, we get $At_0 \approx 10^{-7}$. Substituting this noise amplitude into the error result as shown in Fig. \ref{fig:gammad}(e), the fidelity of the  geometric gates in path 2 is as high as 0.9998. For comparison, the fidelity for the dynamical ones is 0.9995. In Fig. \ref{fig:gammae}, we also show the error results for the systematic noise. We find that when $\alpha\geqslant 1.5$, the geometric gates in both two paths can surpass the dynamical ones. For all $\alpha$ values, the geometric gates in path 1 outperform the ones in path 2.

\begin{figure}
	\includegraphics[width=0.65\columnwidth]{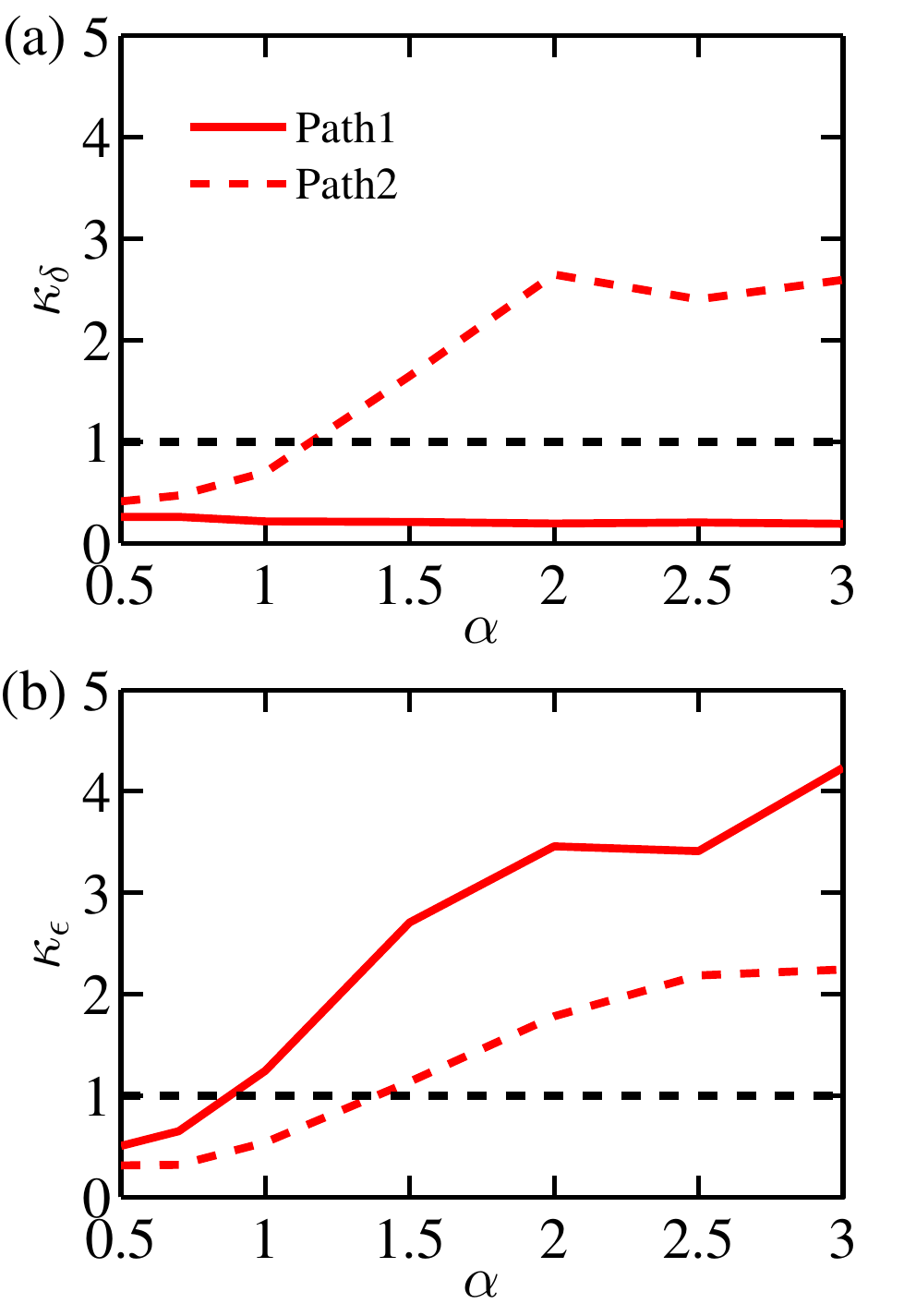}
	\caption{Improvement ratio $\kappa$ vs. exponent $\alpha$ where $\kappa$ is defined as the error of the geometric gates divided by that of the dynamical counterparts . The dashed black line indicates $\kappa=1$.}
	\label{fig:ratio}
\end{figure}

In order to fully reveal the advantage of the geometric gate, we further define the improvement ratio $\kappa$ as the error of the geometric gates divided by that of the dynamical counterparts under the same noise conditions. As we can see in Fig.~\ref{fig:gammad} and \ref{fig:gammae}, the error curves for both the geometric and dynamical gates are almost parallel in the considered noise amplitude region. Such that, the ratio can be well defined. In Fig.~\ref{fig:ratio}, we plot the improvement ratio versus the noise exponent $\alpha$. For the detuning noise, the ratio of the geometric gate in path 1 is always below 1, while that value for the gate in path 2 trends to increase with the exponent $\alpha$. As we can see, the crossing point where the geometric gate starts to outperform the dynamical one is about 1.2. This means the geometric gate can be working well in a rather wide region of $\alpha$. According to the experiments, the $\alpha$ value of the detuning noise can be about either 2.5 \cite{Chan.18, Muhonen.14} or 1 \cite{Chan.18, Yoneda.18, kawakami.16}, which corresponds to the low-frequency and high-frequency domain of the noise spectrum, respectively.  We can also see that, when $\alpha\geqslant2$ the geometric gate can improve the dynamical gate by a factor of 2 or more. This result is a strong evidence to show the advantage of the geometric gate, and is directly relevant to the experiment considering detuning noise is dominant in silicon. While for the systematic noise, the improvement ratio offered by the geometric gates in path 1 can be as high as 4 when $\alpha=3$. And that value for those geometric gates in path 2 can be also larger than 1 when $\alpha\geqslant1.5$. Thus, the geometric gate may be a powerful tool to achieve high-fidelity quantum gates for the experimental noise environment.

\section{Conclusions }\label{sec:conclusion}

We propose the implementation of the universal geometric gate for the silicon spin qubits. By theoretical analysis, we find that, the advantage of the geometric gates over the dynamical couterparts is sensitively depending on the path that are taken for the geometric phase. We also perform randomized benchmarking to quantitatively determine how much improvement the geometric gates can offer. For both the static and $1/f$-type noise model, the fidelities of the geometric gates can be around 99.90\%, or above for both the detuning and systematic noise. For the detuning noise which is dominant in the silicon spin qubits, the proper geometric gates can improve the fidelity of the dynamical gates by a factor of more than 2 in the experimental noise environment. Therefore, our proposal paves a way for implementing the high-fidelity geometric quantum gate for the silicon spin qubits.

\section*{ACKNOWLEDGMENTS}\label{sec:ack}
We thank Bao-Jie Liu for useful discussion. This work was supported by Key-Area Research and Development Program of GuangDong Province  (Grant No. 2018B030326001), the National Natural Science Foundation of China (Grant No. 11905065, 11874156, 11874312), the Project funded by China Postdoctoral Science Foundation (Grant No. 2019M652928),  the National Key R\&D Program of China (Grant No. 2016 YFA0301803), the Research Grants Council of Hong Kong (No. CityU 11303617, CityU 11304018), and the Guangdong Innovative and Entrepreneurial Research Team Program (No. 2016ZT06D348).

\appendix

\setcounter{equation}{0}

\section{Dynamical and geometric phase}\label{appx:phase}

Traditionally, the global phase of the qubit state is assumed to be trivial since there is no physically observable. However, Berry's work revealed that this global phase can not be ignored because it is required for the state to satisfy the schrodinger equation. As stated in the main text, the orthogonal state $\left|\psi_{+}(t)\right\rangle$ can be described as a point on the Bloch sphere. After a cyclic evolution for a period $T$ on the Bloch sphere, it goes back to the starting point and get a global phase factor $f(t)$. Such that, the state satisfying the schrodinger equation associated with $\left|\psi_{+}(t)\right\rangle$ can be denoted as
\begin{equation}
\begin{aligned}
\left|\tilde{\psi}\right\rangle=e^{if(t)}\left|\psi_{+}(t)\right\rangle.
\end{aligned}
\label{app:psi}
\end{equation}
Substitute it into the time-dependent Schr\"{o}dinger equation
%\begin{equation}
%\begin{aligned}
$H(t)|\tilde{\psi}\rangle= i \hbar d |\tilde{\psi}\rangle/d t$,
%\end{aligned}
%\end{equation}
and note that $|\psi_{+}(T)\rangle=|\psi_{+}(0)\rangle$, we have
\begin{equation}
\begin{aligned}
f=\alpha+\beta,
\end{aligned}
\label{tdse}
\end{equation}
where
\begin{equation}
\begin{aligned}
\begin{aligned} \alpha &= -\frac{1}{\hbar} \int_{0}^{T}\langle\widetilde{\psi}|H(t)| \widetilde{\psi}\rangle \mathrm{d} t
\\
&=-\frac{1}{\hbar} \int_{0}^{T}\langle\psi_{+}|H(t)| \psi_{+}\rangle \mathrm{d} t,
\\
\beta &=\int_{0}^{T}\left\langle\widetilde{\psi}\left|\mathrm{i} \frac{\mathrm{d}}{\mathrm{d} t}\right| \widetilde{\psi}\right\rangle \mathrm{d} t.
\end{aligned}
\end{aligned}
\label{app:ab}
\end{equation}
$\alpha$ and $\beta$ are the dynamical and geometric phase, respectively. Next, we show how to cancel out the dynamical phase during the cyclic evolution, i.e. $\alpha=0$. Without loss of generality, a general two-level Hamiltonian has the form as
\begin{equation}
\begin{aligned}
H(t) = \frac{\hbar}{2}\left(\begin{array}{cc}{\Delta(t)} & {\Omega(t) e^{-i \eta(t)}} \\ {\Omega(t) e^{i \eta(t)}} & {-\Delta(t)}\end{array}\right).
\end{aligned}
\label{app:Ht}
\end{equation}
By inserting Eq.~(\ref{app:psi}) and \ref{app:Ht} into Eq. (\ref{tdse}), we can get
\begin{equation}
\begin{aligned}
\dot{\theta}&=\Omega\sin(\eta-\phi)
\\
\dot{\phi}&=\Delta-\Omega\cos(\eta-\phi)\cot\theta,
\end{aligned}
\label{app:thetaphi}
\end{equation}
and further
\begin{equation}
\begin{aligned}
\Omega&=\pm\sqrt{(\Delta-\dot{\phi})^{2}+\dot{\theta}\cot^{2}\theta}\tan\theta
\\
\psi&=\phi-\arctan\left\{\pm\frac{\dot{\theta}\cot\theta}{\Delta-2\dot{\phi}}   \right\},
\end{aligned}
\label{app:thetaphi2}
\end{equation}
Combining Eq.~(\ref{app:Ht}) and \ref{app:thetaphi2} into $\alpha$, we can get
\begin{equation}
\begin{aligned}
\begin{aligned} \alpha &=\int_{0}^{T}\left\{ \frac{\sin^2\theta}{\cos\theta}\dot{\phi}-\frac{\Delta}{\cos\theta}\right\}dt
\end{aligned}
\label{eq:deta}
\end{aligned}
\end{equation}
Now, it is clear that, when $\Delta=0$, if we take $\dot{\phi}=0$, then the dynamical phase can be canceled out. This implies the state $\left|\psi_{+}(t)\right\rangle$ would evolve along the geodesic line. According to Eq.~(\ref{app:thetaphi}), we further have $\eta-\phi=\pm \pi/2$.

%\section{Clifford gates for randomized benchmarking}\label{appx:retable}

%\bibliography{refs_GQG}

%

\end{document}